\documentclass[12pt]{article}
\usepackage{amsmath}
\usepackage{graphicx}
\begin{document}
\baselineskip=18 pt
\begin{center}
{\large{\bf  Comment on ``An axially symmetric spacetime with causality violation [Phys. Scr. {\bf 96} (07), 075208 (2021)], arXiv:2004.13769[gr-qc]" }}
\end{center}

\vspace{.5cm}

\begin{center}
{\bf Faizuddin Ahmed}\footnote{\bf faizuddinahmed15@gmail.com ; faiz4U.enter@rediffmail.com}\\
{\bf National Academy Gauripur, Assam, India, 783331}
\end{center}

\vspace{.5cm}

\begin{abstract}

Recently, an axially symmetric spacetime with causality violation is appeared in Ref. \cite{BB}. There, author presented a type III metric with vanishing expansion, shear and twist. The matter-energy represents a pure radiation field with a negative cosmological constant. The spacetime is asymptotically anti-de Sitter space in the radial direction. Actually, this work with different form of the metric was already published in Ref. \cite{FA}. Therefore, the current paper Ref. \cite{BB} is a duplicate one of the previous work.

\end{abstract}

\vspace{0.5cm}

Recently, an axially symmetric metric with causality violation is appeared in Ref. \cite{BB}. The metric in cylindrical coordinates $(t, r, \phi, z)$ is given by
\begin{equation}
ds^2=\frac{dr^2}{\alpha^2\,r^2}+r^2\,(dz^2-2\,dt\,d\phi-t\,d\phi^2)+\frac{\beta\,z}{r^2}\,dr\,d\phi,
\label{1}
\end{equation}
where $\alpha >0, \beta>0$ are non-zero real numbers. The coordinates are in the ranges $
-\infty < t < \infty$, $0 \leq r < \infty$, $-\infty < z < \infty$ and $\phi$ is a periodic. 

The scalar curvature invariant constructed from the Riemann tensor are 
\begin{equation}
R_{\mu\nu\rho\sigma}\,R^{\mu\nu\rho\sigma}=24\,\alpha^4.
\label{2}
\end{equation}
with the scalar curvature
\begin{equation}
R=R^{\mu}_{\,\mu}=-12\,\alpha^2.
\label{3}
\end{equation}
The space-time (\ref{1}) satisfy the Field Equations 
\begin{equation}
R^{\mu\nu}=\Lambda\,g^{\mu\nu}+\rho\,k^{\mu}\,k^{\nu}
\label{EFE}
\end{equation}
with the stress-energy tensor the pure radiation field. The radiation energy density with a negative Cosmology constant is given by  
\begin{equation}
\Lambda=-3\,\alpha^2\quad,\quad \rho=\frac{\alpha^2\,\beta^2}{8\,r^8},
\label{4}
\end{equation}
where the null vector for the metric (\ref{1})
\begin{equation}
k^{\mu}=(1,0,0,0)
\label{5}
\end{equation}
satisfies null condition $k_{\mu}\,k^{\mu}=0$. Therefore, the energy-density of radiation field satisfies the null energy condition (NEC).

One can easily show that the null vector (\ref{5}) is a covariantly constant null vector field, that is,
\begin{equation}
k_{\mu\,;\,\nu}=0
\label{6}
\end{equation}
To classify the metric (\ref{1}) according to the Petrov classification scheme, author constructed a set of tetrad vectors $({\bf k, l, m, {\bar m}})$. Using the set of tetrad vectors, one can calculate the five Weyl scalars and these are 
\begin{equation}
\Psi_0=\Psi_1=0=\Psi_2\quad,\quad \Psi_3=\frac{i\,\alpha^2\,\beta}{4\,\sqrt{2}\,r^2}\quad,\quad \Psi_4=-\frac{\alpha\,\beta\,(i+2\,r\,z\,\alpha)}{8\,r^2}.
\label{7}
\end{equation}
Therefore the metric represented by (\ref{1}) is of type III in the Petrov classification scheme Ref. \cite{Steph}. 

One can easily show that the metric (\ref{1}) is non-expanding, non-twisting and shear-free which are as follows
\begin{eqnarray}
&&\Theta=\frac{1}{2}\,k^{\mu}_{\,;\mu}=0,\nonumber\\
&&\omega=\sqrt{\frac{1}{2}\,k_{[\mu;\nu]}\,k^{\mu;\nu}}=0,\nonumber\\
&&|\sigma|=\sqrt{\frac{1}{2}\,k_{(\mu;\nu)}\,k^{\mu;\nu}-\Theta^2}=0.
\label{condition}
\end{eqnarray}

The study space-time admits closed time-like curves (CTCs). Consider an azimuthal curve $\gamma$ defined by \{$ t=const.$, $r=const.$, $z=const. $\}, and $\phi$ is a periodic coordinate. From the metric (\ref{1}), we have
\begin{equation}
ds^2=-t\,r^2\,d\phi^2.
\label{8}
\end{equation}
The integral curve with $(t, r, z)$ fixed is time-like provided $t>0$ and thus closed time-like curves are formed at an instant of time Ref. \cite{EPJC}.

The original metric with different form was already obtained by other author in Ref. \cite{FA}. The original metric in cylindrical coordinate is given by
\begin{equation}
ds^2=4\,r^2\,dr^2+e^{2\,\alpha\,r^2}\,(dz^2-t\,d\phi^2-2\,dt\,d\phi)+4\,\beta\,z\,r\,e^{-\alpha\,r^2}\,dr\,d\phi,
\label{9}
\end{equation}
where $\alpha, \beta$ are non-zero real numbers. We do the following transformation
\begin{equation}
r \to \sqrt{\varrho}
\label{10}
\end{equation}
into the original metric (\ref{9}), one will get
\begin{equation}
ds^2=d\varrho^2+e^{2\,\alpha\,\varrho}\,(dz^2-t\,d\phi^2-2\,dt\,d\phi)+2\,\beta\,z\,e^{-\alpha\,\varrho}\,d\varrho\,d\phi.
\label{11}
\end{equation}
Finally doing another transformation as follows
\begin{equation}
\varrho \to \frac{1}{\alpha}\,\mbox{ln} r'
\label{12}
\end{equation}
into the metric (\ref{11}), we get
\begin{equation}
ds^2=\frac{dr'^2}{\alpha^2\,r'^2}+r'^2\,(dz^2-2\,dt\,d\phi-t\,d\phi^2)+\frac{2\,\beta\,z}{\alpha\,r'^2}\,dr'\,d\phi,
\label{13}
\end{equation}
which is very similar to the metric (\ref{1}). Therefore, the results obtained in the current paper Ref. \cite{BB} are same with the previous work Ref. \cite{FA}. So, the current paper Ref. \cite{BB} is a duplicate one of the previous work.


\begin{thebibliography}{99}

\bibitem{BB} Bidyut Bikash Hazarika, {\tt An axially symmetric spacetime with causality violation}, Phys. Scr. {\bf 96} (07), 075208 (2021), arXiv : 2004.13769v1 [gr-qc].

\bibitem{Steph} H. Stephani, D. Kramer, M. MacCallum, C. Hoenselaers and E. Herlt, {\tt Exact Solutions to Einstein’s Field Equations}, Cambridge University Press, Cambridge (2005). 

\bibitem{EPJC} F. Ahmed, {\tt A type N radiation field solution with $\Lambda <0$ in a curved space-time and closed time-like curves}, Eur. Phys. J. C {\bf 78}, 385 (2018).

\bibitem{FA} F. Ahmed, {\tt Type III Spacetime with Closed Timelike Curves}, PROGRESS IN PHYSICS (PiP) Volume {\bf 12}, Issue 4 (October), pp. 329-331 (2016), http://www.ptep-online.com.

\end{thebibliography}
\end{document}